# Anisotropic quantum transport in monolayer graphene in the presence of Rashba spin-orbit coupling

Kobra Hasanirokh , Hakimeh Mohammadpour , Arash Phirouznia

Azarbaijan Shahid Madani University, 53714-161, Tabriz, Iran

**Abstract:** *We have studied* spin-depend electron tunnelling through the Rashba barrier in a monolayer graphene lattices. The transfer matrix method, have been employed to obtain the spin dependent transport properties of the chiral particles. It is shown that graphene sheets in presence of Rashba spin-orbit barrier will act as an electron spin- inverter.

## 1. Introduction

Graphene with quasi relativistic energy spectrum of Dirac fermions[1,2] and unconventional quantum Hall effect (QHE)[3-5] has attracted more and more attention. Experimental studies in graphene have shown that the electron-spin relaxation length is very long (1μm at room temperature)[6,7], so graphene is a promising candidate for spintronic devices and prospective applications in nano-elecronics.

Electronics of the spin-dependent electron transport has been studied extensively in graphene[8,9]. Applications in spintronics depend on the control of spin-orbit (SO) coupling[10-12]. Spin- orbit interaction in graphene, can manipulate the spin of electrons. This interaction comprises of two different type i.e. intrinsic and extrinsic (Rashba) couplings[13,14]. First one is induced by carbon intra-atomic spin-orbit interaction and can open a gap in graphene energy dispersion This interaction can convert graphene in to a topological insulator and induces some other interesting effects like fractional spin Hall effect[15]. Intrinsic SOI is very weak in a free plane graphene. Theoretical calculations in graphene have been shown that the strength of the extrinsic spin-orbit coupling can be remarkably higher than intrinsic spin-orbit interaction[16-18] . The Rashba spin-orbit interaction that arises from the structure inversion asymmetry (SIA), is introduced by a substrate surface or an external field. [19]

The ferromagnetic layer in magnetic tunnel junctions (MTJ)[20] splits the energies of the electrons in to two sub-bands. Therefore this splitting leads to spin-dependent conductance and characterizes the spin-polarized current that passing through the junctions. Manipulating the spin degree of freedom is of high importance in the field spintronics and one of the main purposes of spin manipulation technique is the electron spin inversion. Spin inversion in emiconductors has been studied vastly, but there are few studies in nano-scale semiconductor.[21,22] Busl and Platro have shown a triple quantum dot in dc and ac magnetic fields, by tuning in gate voltages of two dots, will act as spin- inverter. In Ref 22, it has been shown that graphene sheets in the presence of Rashba sbin-orbit barrier will act as an electron spin- inverter.

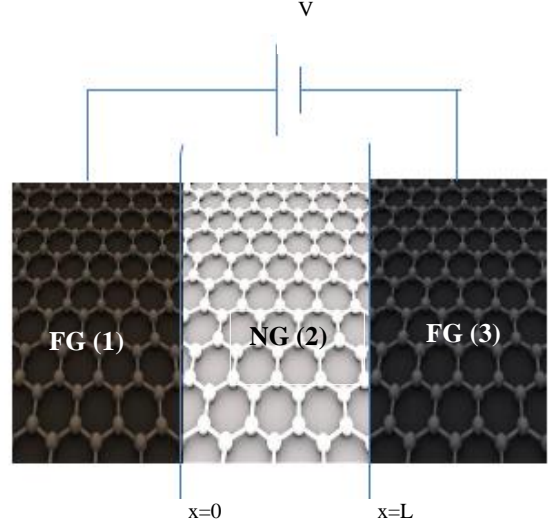

Fig. 1. Schematic illustration of the Rashba spin-orbit barrier with height $\lambda_R$ and length L.

## 2. Model and approach

In this work, spin transport of the carriers in two sub-bands have been studied in which the Weyl-Dirac equation describes the dynamics of the carriers in the graphene. The system considered here is a ferromagnetic-Rashba barrier-ferromagnetic ( FG/NG/FG ) junction located in the x-y plane, that x is normal to the interfaces between the various types of graphene. NG denotes a nonmagnetic graphene layer which is affected by Rashba spin-orbit interaction generated by a gate voltage of potential V placed above the sheet, as shown in Fig1. Dimensions of graphene sheet are assumed to be large enough, so we can ignore edge effects. that $\tau_z = \pm 1$ indicate the states on the A and B sub-lattices of graphene while $\sigma_z = \pm 1$ corresponds to the up and down spins, respectively.

The Hamiltonian in regions (1) and (3) is

$$H = H_0 + H_I + H_\sigma \quad (1)$$

where $H_0$ is the Dirac Hamiltonian for massless fermions:

$$\hat{H}_0 = -i\hbar v_f \psi^t (\sigma_x \tau \partial_x + \sigma_y \partial_y)\psi \quad (2)$$

$\hbar$ is the Planck constant, $v_F$ denotes the Fermi velocity in graphene and σ and τ are Pauli matrices. $\tau_z = \pm 1$ describe the states on the A and B sub lattices in graphene and $\sigma_z = \pm 1$ indicates up and down spins .
$H_I$ is intrinsic (SO) interaction that is small and we can ignore it.
We have considered regions (1) and (3) as two different ferromagnetic regions, so we denote the $H_\sigma$ in region (1) as

$$H_\sigma = h\sigma, \quad (3)$$

and in region (3) as

$$H_\sigma = h'\sigma. \quad (4)$$

The Hamiltonian in the central region with Rashba interaction is
$$H = H_0 + H_R \quad (5)$$

$H_R$ is the Rashba (SO) interaction, and can be written as[23,24]

$$\hat{H}_R = \lambda_R / 2 \psi^t (\sigma_y s_x - \sigma_x \tau_z s_y) \psi \quad (6)$$

Where $\lambda_R$ is the strength of Rashba (SO) interaction and $s_i$ denotes Pauli matrix representing the spin of electron.

The regions $x \leq 0$ and $x \geq L$ taken to be left and right ferromagnetic graphene (FG) layers, respectively, are shown in Fig1.
We have considered a typical electron with the given energy E and up spin orientation, propagating from the left of nano-scale single sheet of graphene to the Rashba region.
The wave functions of the electrons exhibit the chiral properties of the graphene. The solutions of the Dirac equations (1) and (5), will be in the following spinor form:

$$\psi(x \leq 0) = \exp(ikx\cos\varphi)\begin{pmatrix} e^{-i\varphi/2} \\ e^{i\varphi/2} \\ 0 \\ 0 \end{pmatrix} + r\exp(-ikx\cos\varphi)$$

$$\begin{bmatrix} e^{-i(\pi-\varphi)/2} \\ e^{i(\pi-\varphi)/2} \\ 0 \\ 0 \end{bmatrix} + r'\exp(-ik'x\cos\varphi')\begin{bmatrix} 0 \\ 0 \\ e^{-i(\pi-\varphi')/2} \\ e^{i(\pi-\varphi')/2} \end{bmatrix}$$

(7)

That
$$k' = \frac{E-h}{\hbar v_f}, k = \frac{E+h}{\hbar v_f}, \varphi' = \sin^{-1}(k\tan\varphi/k').$$

We assumed that the incident electrons made the angles $\varphi$ (for spin up) and $\varphi'$ (for spin down) in region 1 and made angles $\varphi_1$, $\varphi_2$, $\varphi_3$ and $\varphi_4$, in region 2 and also made $\overline{\varphi}$ (for spin up) and $\overline{\varphi}'$ (for spin down) angles in region 3.

$$\psi(0 \leq x \leq L)$$

$$= A\exp(ik_1 x \cos\varphi_1) \begin{bmatrix} \frac{\hbar v_f k_1}{E} \\ 1 \\ \frac{E - \frac{\hbar^2 v_f^2 k_1^2}{E}}{i\lambda_R} \\ \frac{\hbar v_f k_1}{i\lambda_R}\left(E - \frac{\hbar^2 v_f^2 k_1^2}{E}\right)e^{i\varphi_1} \end{bmatrix}$$

$$+ B\exp(ik_2 x \cos\varphi_2) \begin{bmatrix} \frac{\hbar v_f k_2}{E} \\ 1 \\ \frac{E - \frac{\hbar^2 v_f^2 k_2^2}{E}}{i\lambda_R} \\ \frac{\hbar v_f k_2}{i\lambda_R}\left(E - \frac{\hbar^2 v_f^2 k_2^2}{E}\right)e^{i\varphi_2} \end{bmatrix}$$

$$+ C\exp(ik_3 x \cos\varphi_3) \begin{bmatrix} \frac{\hbar v_f k_3}{E} \\ 1 \\ \frac{E - \frac{\hbar^2 v_f^2 k_3^2}{E}}{i\lambda_R} \\ \frac{\hbar v_f k_3}{i\lambda_R}\left(E - \frac{\hbar^2 v_f^2 k_3^2}{E}\right)e^{i\varphi_3} \end{bmatrix}$$

$$+ D\exp(ik_4 x \cos\varphi_4) \begin{bmatrix} \frac{\hbar v_f k_4}{E} \\ 1 \\ \frac{E - \frac{\hbar^2 v_f^2 k_4^2}{E}}{i\lambda_R} \\ \frac{\hbar v_f k_4}{i\lambda_R}(E - \frac{\hbar^2 v_f^2 k_4^2}{E})e^{i\varphi_4} \end{bmatrix}$$

(8)

Where

$$\varphi_1 = \tan^{-1}(k\tan\varphi/k_1)$$
$$\varphi_2 = \tan^{-1}(k\tan\varphi/k_2). \qquad (9)$$
$$\varphi_3 = \pi - \varphi_1, \varphi_4 = \pi - \varphi_2$$

r, t, r ', t' are the scattering amplitudes in ferromagnetic regions (1) and (3) and A, B, C, D are the coefficients of the electronic wave function in the Rashba region.

$$\psi(x \geq L) = t\exp(\bar{k}x\cos\bar{\varphi})\begin{pmatrix} e^{-i\bar{\phi}/2}/\sqrt{\cos\bar{\varphi}} \\ e^{i\bar{\phi}/2}/\sqrt{\cos\bar{\varphi}} \\ 0 \\ 0 \end{pmatrix} + t'\exp(-i\bar{k}'x\cos\bar{\varphi}')\begin{bmatrix} 0 \\ 0 \\ e^{-i\bar{\varphi}'/2}/\sqrt{\cos\bar{\varphi}'} \\ e^{i\bar{\varphi}'/2}/\sqrt{\cos\bar{\varphi}'} \end{bmatrix}$$
$$(10)$$

where
$$k_1 = \frac{\sqrt{(E+V)^2 - \lambda_R(E+V)}}{\hbar v_f},$$
$$k_2 = \frac{\sqrt{(E+V)^2 + \lambda_R(E+V)}}{\hbar v_f},$$
$$k_3 = -k_1, k_4 = -k_2$$
$$(11)$$

where

$$\bar{k}' = \frac{E - h'}{\hbar v_f}, \quad \bar{k} = \frac{E + h'}{\hbar v_f}$$

The factors $\frac{1}{\sqrt{\cos\bar{\varphi}}}$, $\frac{1}{\sqrt{\cos\bar{\varphi}'}}$ in above equations ensure that all of the four states carry the same particle current.

From the conservation of momentum along the y-axis, we have

$$k\sin\varphi = k_1\sin\varphi_1 = \cdots \qquad (12)$$

At the interfaces, the wave functions have to be continuous[25] ,so

$$\psi(0^-) = \psi(0^+), \qquad \psi(L^-) = \psi(L^+)$$
$$(13)$$

We obtain the amplitudes by applying the boundary conditions at the interfaces. Where theses amplitudes characterizes the transport properties of the system.

### 3. Results

A numerical study of spin-inversion properties in FG/N'/FG is presented in this section using Eqs.(1) and (5). The Fermi energy is taken to be $E_f = 1meV$ in all numerical calculations. All energies is written in term of $E_f$.
Here, the incoming electron is assumed to be polarized with up spin state, the length of Rashba spin-orbit is taken to be L=2nm.

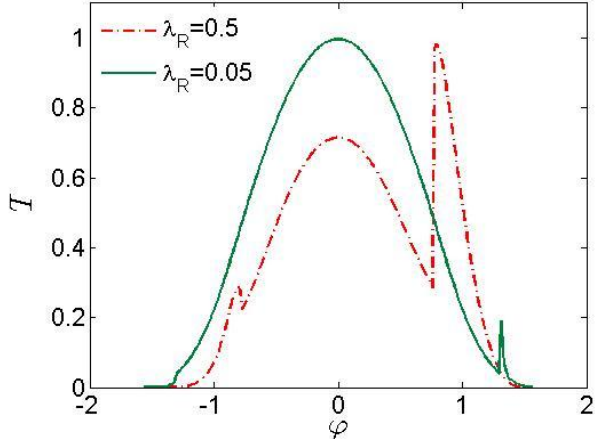

Fig2. T ($T_{\uparrow\to\uparrow}$) as a function of electron incident angle ϕ ($h_1=h_2=0.01$).

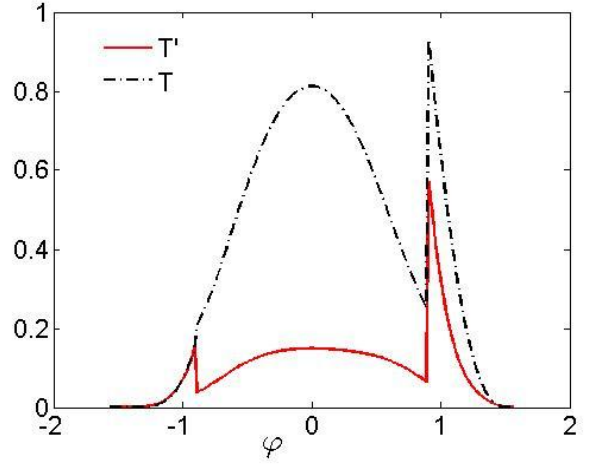

Fig3. T ' ($T_{\uparrow\to\downarrow}$) and T ($T_{\uparrow\to\uparrow}$) as function of incident angle φ ($h_1=h_2=-0.01$, V=0, $\lambda$=0.4).

The difference between these curves clearly demonstrates the importance of the Rashba coupling.

The spin-dependent transmission coefficients with spin inversion ($T_{\uparrow\to\downarrow}$) and without spin-inversion ($T_{\uparrow\to\uparrow}$) as a function of electron incident angle ϕ are shown in Fig3. Increasing the value of the Rashba barrier, can disturb the symmetry of transmission coefficients of both spin-flip and non spin-flip transports. This is due to the fact that in the presence of the Rashba interaction the Hamiltonian is not invariant under the $\sigma_y \to -\sigma_y$ transformation. Therefore one can easily obtain that $T_{\sigma\to\sigma}(\varphi) \neq T_{\sigma\to\sigma}(-\varphi)$. This anisotropy which was induced by Rashba interaction indicates that in the present system, some directions can be considered as a spin selective path in which the transmission probability is higher than other directions. This interesting feature could be employed directional spin filtering devices.

Note that the Rashba SO interaction acts like an effective magnetic field oriented in direction of y axis when the carrier is moving along the x axis. The spin of electrons moving in direction of x axis rotates in y-direction due to this effective magnetic field as shown in Fig4.

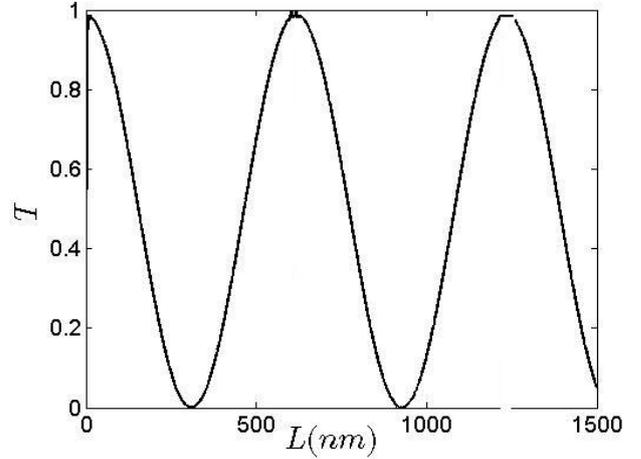

Fig4. T ($T_{\uparrow\to\uparrow}$) as a function of L ($h_1=h_2=0.15$, V=0, $\lambda$=0.01, φ= π/20).

Meanwhile by increasing the gate voltage up to V=200 we can obtain almost isotropic transmission probability with and without spin-inversion transmissions.

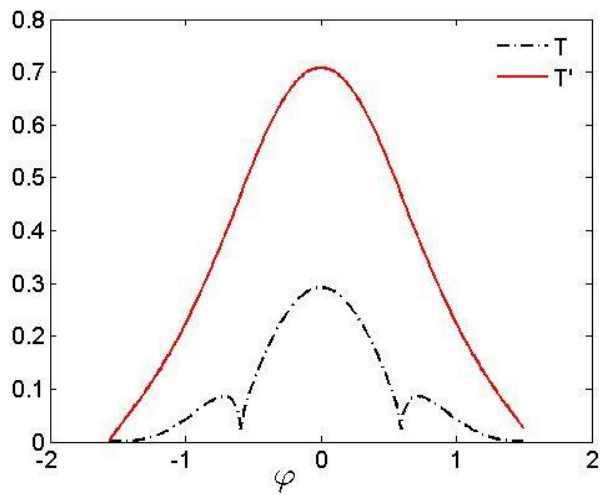

Fig 5. T ' ($T_{\uparrow\rightarrow\downarrow}$) and T ($T_{\uparrow\rightarrow\uparrow}$) as function of incident angle φ ( $h_1$ = - 0.1, $h_2$ = - 0.5, V =200, λ =1).

In Fig 6, $T_{\uparrow\rightarrow\downarrow}$ has been depicted as a function of the Rashba interaction. Increment of the absolute value of $h_2$ increases the amplitude of oscillations. This figure clearly shows that transmission of the system, with respect to the $h_2$, can be controlled by Rashba coupling.

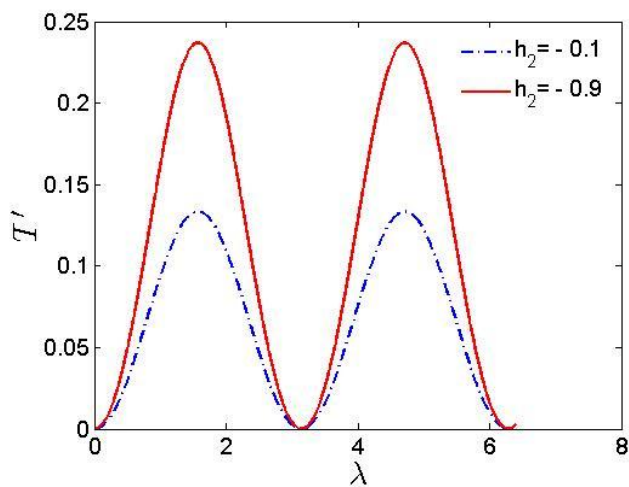

Fig 6. T ' ($T_{\uparrow\rightarrow\downarrow}$) as function of Rashba coupling λ ( $h_1$ = -0.1 , V =200, φ= π/3).

$T_{\uparrow\rightarrow\downarrow}$ has been shown as a function of the L in Fig.7 . Transport properties can be effectively changed by the $h_2$ in which the amplitude of oscillations in transmission coefficients are increased by increment rate of the absolute value of $h_2$.

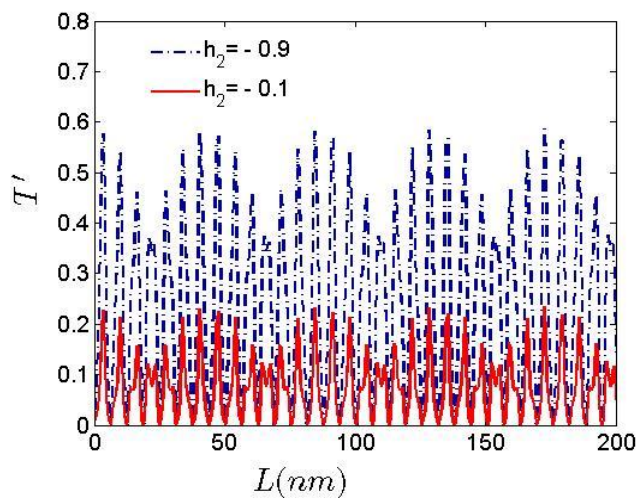

Fig 7. T ' ($T_{\uparrow\rightarrow\downarrow}$) as function of L ( $h_1$ = - 0.01, λ = 1, V=200 , φ= π/3).

## 4. Conclusion

In present work the anisotropic quantum transport in FG/NG/FG junction have been studied. Rashba SO interaction can control the symmetry of transport and when the carrier is the x direction, Rashba acts like an effective magnetic field oriented in the along of y axis. Results also show that beside the critical role played by Rashba SOI effect, other parameters, i.e. U, $h_2$ and L can also effectively control the transport properties.

## 5. Acknowledgement

This research has been supported by Azarbaijan Shahid Madani university